# Direct measurement of the acoustic waves generated by femtosecond filaments in air

J. K. Wahlstrand, N. Jhajj, E. W. Rosenthal, S. Zahedpour, and H. M. Milchberg
*Institute for Research in Electronics and Applied Physics*
*University of Maryland, College Park, MD 20742*

**Abstract**

We present direct measurements of the gas acoustic dynamics following interaction of spatial single- and multi-mode 50 fs, 800 nm pulses in air at 10 Hz and 1 kHz repetition rates. Results are in excellent agreement with hydrodynamic simulations. Under no conditions for single filaments do we find on-axis enhancement of gas density; this occurs only with multi-filaments. We also investigate the propagation of probe beams in the gas density profile induced by a single extended filament. We find that light trapping in the expanding annular acoustic wave can create the impression of on-axis guiding in a limited temporal window.

The generation of acoustic waves in gases by intense ultrashort laser filaments has been noted in various contexts [1-4]. The physical mechanism behind sound generation is the sudden deposition of energy in the gas, which produces a localized pressure spike that relaxes by producing an outwardly propagating acoustic wave [5,6]. A region of increased temperature and lower density is left behind that then slowly dissipates by thermal diffusion over millisecond time scales. The long-lived thermal depressions produced by short pulses were recently measured interferometrically [7], but those measurements did not have the sub-microsecond time resolution required to measure the full evolution of the acoustic response. Levi *et al*. recently used a delayed 150 ns pulse to probe the axially extended gas density perturbation produced by an ultrashort filamenting pulse in air [8]. They argued that there must be a positive gas density perturbation on axis with a microsecond lifetime that supports optical guiding.

Here, we present interferometric measurements of the acoustic response induced by filamenting single- and multi-mode beams, at repetition rates of 10 Hz and 1 kHz, and we compare our measurements to hydrodynamic simulations. We also perform simulations of the propagation of light in post-filament gas density profiles. For single filaments, at no delay do we find on-axis enhanced gas refractive index profiles that can support guided modes; such profiles are produced only by multi-filaments.

Gas dynamics after deposition of energy by an ultrashort optical pulse has been discussed previously [5-7,9,10]; we briefly review the subject here. A weakly ionized plasma created through field ionization of molecules recombines over a timescale of ~10 ns and repartitions its energy into the translational and rotational degrees of freedom of the neutral gas [7]. Non-resonant two-photon rotational Raman absorption also contributes to heating for pulses shorter than ~150 fs in air [3,11]. Because of the increased energy density (and thus pressure), an acoustic wave is launched radially, and after ~1 μs the gas reaches pressure equilibrium as a heated, low density central region surrounded by room temperature gas at ambient density, with further relaxation by thermal diffusion over milliseconds [7].

We simulate the gas dynamics using a one-dimensional radial Lagrangian hydrodynamic code that solves the partial differential equations for mass, momentum and energy conservation, and for thermal diffusion. For typical filament-deposited energy density in a gas, radiation from the heated medium contributes negligibly to the overall energy balance and can be neglected, as can be verified by a simple black-body estimate. Details of the model are described in [7]. As we are interested in acoustic timescales >10 ns, we can simplify the initial conditions as follows. Immediately after filament generation, absorbed laser energy is invested in plasma generation and coherent excitation of molecular rotational states [11]. As plasma recombination occurs by ~10 ns and rotational coherence is lost through collisions over hundreds of picoseconds, we assume repartitioning of the initial energy deposition to the translational and rotational degrees of freedom of the neutral gas [7].

Simulation results assuming a 30 μm FWHM Gaussian heat source of energy density 30 mJ/cm$^3$, are shown in Fig. 1a. The refractive index of air at standard atmospheric temperature and pressure, $n_0-1 = 2.7 \times 10^{-4}$ [12], was used to calculate the refractive index change from the gas density. The single cycle acoustic wave propagates outward at the speed of sound, ~340 m/s in air [13], and the amplitude of the density perturbation falls as $\sim r^{-1/2}$. The central density depression rapidly reaches its maximum depth by ~100 ns and then changes very little over microsecond timescales after the acoustic wave is launched.

Filaments were generated with 50 fs, 800 nm Ti:Sapphire pulses at 10 Hz to investigate single shot gas response and at 1 kHz to probe for a possible dependence on repetition rate [7,10]. A 532 nm, 7 ns laser probe pulse timed to the Ti:Sapphire laser is used in a counter-propagating geometry. The optical setup is similar to that in [7] and [8], with the important exception that here and in [7] we perform imaging interferometry for direct and unequivocal measurement of the evolving refractive index profile. We use folded wavefront interferometry [7,9,10], which measures the 2D phase shift $\Delta\phi(x,y)$ of the probe beam, from which the refractive index perturbation $\Delta n$ is extracted. Timing jitter between the two laser systems is <10 ns, which is unimportant for the longer timescales explored in this experiment. For a reliable measurement of $\Delta\phi(x,y)$, it is critically important to keep the interaction length of the probe short enough that refraction from the gas perturbation negligibly distorts the probe [14]. This is achieved by limiting the filament length to ~2mm by focusing the pump laser beam relatively tightly at $f/30$. In the absence of refractive distortion, the change in probe phase due to the index perturbation $\Delta n$ is $\Delta\phi(x,y) = k\int\Delta n(x,y,z)\,dz \approx k\,\Delta n(x,y,0)\,L_{\text{eff}}$ where $k$ is the vacuum wavenumber of the probe, $z=0$ is the pump beam waist location, and $L_{\text{eff}} = 2$ mm.

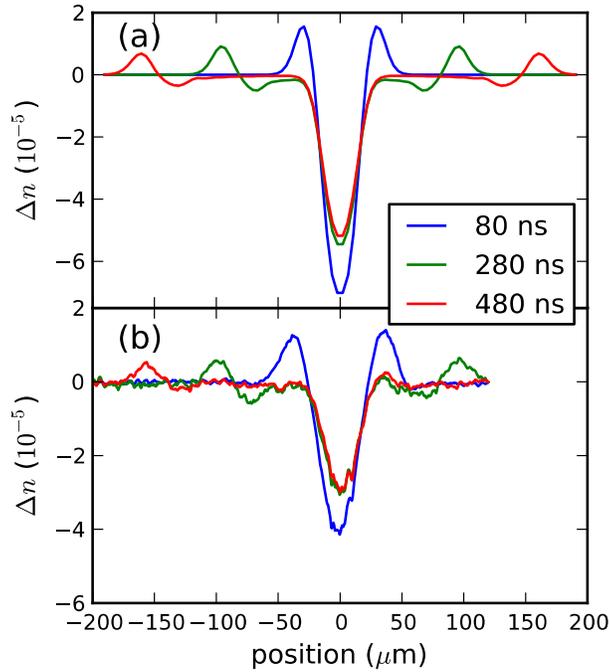

**Figure 1.** Acoustic gas dynamics following an ionizing pulse in air. The refractive index change Δ*n* is shown at time *T* after the passage of the pulse. Blue: *T* = 80 ns; green: *T* = 280 ns; red: *T* = 480 ns. (a) Hydrocode simulation. The calculated Δ*n* is plotted as a function of transverse coordinate. (b) Sequence of Δ*n* profile lineouts obtained by interferometry of a 2 mm long plasma at 10 Hz [9].

The measured refractive index change induced by the 10 Hz laser with a 96 µJ pulse (vacuum peak intensity ~ $4 \times 10^{14}$ W/cm$^2$) is shown in Fig. 1b for a few delays following excitation. The density depression along the optical axis is clearly seen, as is the outwardly propagating single cycle acoustic wave. Excellent agreement is found between the experiment (Fig. 1b) and the calculation (Fig. 1a).

We repeated our measurements with the 1 kHz laser with pulse energy 60 µJ (vacuum peak intensity~$2.5 \times 10^{14}$ W/cm$^2$) and the same focusing geometry. The measured refractive index profile is shown in Fig. 2. We found no qualitative differences between the 10 Hz and 1 kHz cases apart from the broad and shallow density hole left over from the previous pulse, where the depth of the hole at 1 ms is a few percent of ambient density [7]. Such a hole can negatively lens an optical pulse over an extended region [7,10], but has negligible effect on a tightly focused beam. We also observe that the hydrodynamic evolution of the gas initiated by individual pulses in the 1 kHz train is essentially unaffected by the presence of the preexisting density hole.

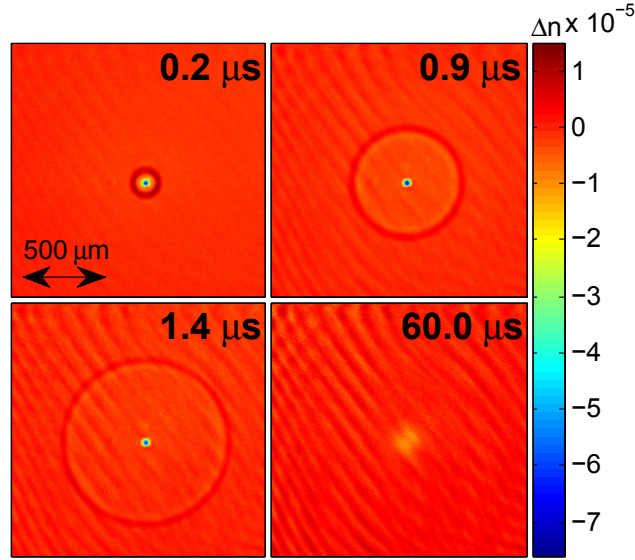

**Figure 2**. Measured refractive index profiles induced by 60 µJ pump pulses at 1 kHz. The delays shown are with respect to passage of the pump pulse. The thermal gas density hole left between pulses is seen in the 60 µs delay panel.

Next, we generated double and octuple filaments using $TEM_{01}$ and Laguerre-Gaussian $LG_{04}$ modes at the focus. For a $TEM_{01}$ focal mode, the pre-focused beam was passed through a half-pellicle angled so as to phase shift one half the beam by $\pi$ with respect to the other half. For the $LG_{04}$ focus, with eight azimuthal intensity lobes, we use normal incidence reflection of the pre-focused beam from an eight segment mirror with alternating triangular segments recessed by $\lambda/4 \sim 200$ nm or $\pi/2$. Using comparable methods, other groups have produced multi-lobed beams for filament generation [15,16]. Images of the acoustic response to multi-lobed modes are shown in Fig. 3. Figure 3(a) and (b) show the measured response at 0.2 µs and 0.5 µs for a $TEM_{01}$ mode. Figure 3(c) and (d) show simulations assuming that the gas dynamics is linear, so that two radially offset single filament results can be added. This is reasonable since the peak relative amplitude of the acoustic wave, $\Delta\rho/\rho \sim 0.05$, is small. The approximation is borne out by the good agreement with the measurements.

The measured acoustic response from the $LG_{04}$ mode is shown in Fig. 3(e) and (f). The low intensity 8-lobe mode is shown in the figure inset. The gas response to the $LG_{04}$ focus was simulated assuming a continuous ring heat source at $t=0$ with a Gaussian cross section at the ring location. The results, shown in Fig 3(g) and (h), are in excellent agreement with the experimental results. The ring does a very effective job of simulating the merged acoustic response to closely spaced azimuthal lobes. As seen in the experiment and simulation, at longer delays (Fig. 3(f) and (h)), two single cycle sound waves propagate away from the beam axis. This is because the lobes (or ring) launch both inward- and outward-directed acoustic pulses. The inward-directed annular acoustic wave collides with itself on axis and produces a very strong density increase of ~30% (Fig. 3(e) and (g)). It then passes through the axis and re-emerges as a wave trailing the originally outward-directed wave. Movies of the gas evolution induced by single- and multi-filaments –both experiment and simulation- can be viewed at [17].

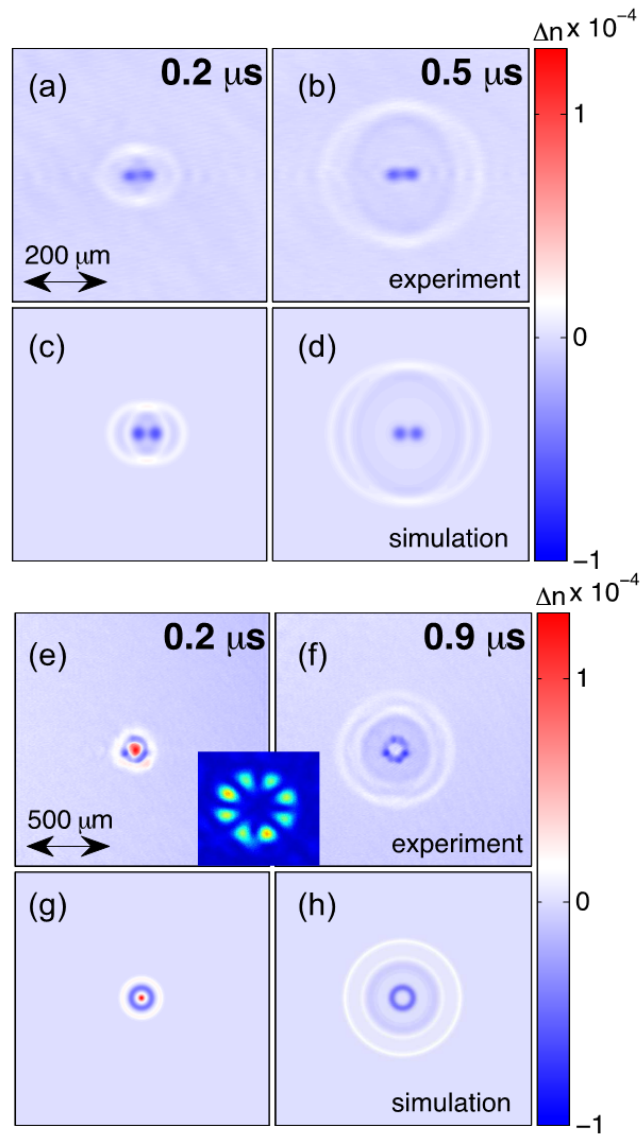

**Figure 3.** Measured and simulated air refractive index profiles induced by a TEM$_{01}$ focus (vacuum peak lobe intensity~$2.5\times10^{14}$ W/cm$^2$) ((a)-(d)) and an LG$_{04}$ focus (vacuum peak lobe intensity~$3\times10^{14}$ W/cm$^2$) ((e)-(h)). The inset between (e) and (f) shows an expanded view of the focused LG$_{04}$ beam mode at low intensity. The simulations in (g) and (h) assume a continuous ring heat source at $t=0$ with a Gaussian cross section at the ring location.

We have generated local density enhancements on-axis only in the case of gas dynamics initiated by a multi-lobed focus, as seen in Fig. 3 for LG$_{04}$ and TEM$_{01}$ modes, and with TEM$_{11}$ modes (see [17]). As described above, in those cases, acoustic waves are launched from the lobe locations and superpose near the axis. However, for single filaments generated by a wide range of pump intensities and focal spot sizes, we always find that the central density (and refractive index) change is negative with respect to the surrounding gas. Such a profile would not be expected to guide an on-axis beam. But the results of Levi *et al.* [8] show apparent on-axis guiding of a 150 ns probe pulse at ~ 1 µs delay in their post-single-filament gas density profile. What can explain this? In Levi *et al.*, the filament length is ~300 mm, while for our interferometry measurements it is intentionally kept short at $L_{eff}$ ≈ 2 mm. There is no reason to believe that the radial hydrodynamics are significantly different in the two cases. Our simulation assumes an infinitely long heat source and it matches our short filament measurements extremely well. However, what remains to be considered are the details of probe pulse propagation in the extended gas density profile produced by a long filament.

To investigate this issue, we performed experiments and simulations for an extended interaction length; results are shown in Fig. 4. A ~20 cm filament was generated by a 10 Hz, 2.8 mJ, 50 fs, 800 nm pulse focused at f/200. For a better understanding of a long filament's effect on a probe beam, one must know the axial dependence of the induced refractive index perturbation. To characterize the axial profile of the energy deposited by the filament, we used a sonographic technique [1]. A rail-mounted microphone is positioned a few mm away from the filament and scanned along it. The amplitude of the leading feature of the sound trace (see inset of Fig. 4a) is plotted as a function of axial position in Fig. 4. The sound amplitude is proportional to the laser energy deposited.

This information was fed into a beam propagation method simulation [18], the results of which are shown in Fig. 4b, 4c, and 4d for probe injection delays of 0.4 µs, 0.8 µs, and 1.2 µs, respectively. Here, the simulated gas density structure was built by weighting the results of the hydrocode by the axial slice-by-slice relative laser energy deposition measured in Fig. 4a. The structure is stationary on the timescale of the 7ns probe pulse. Note the enhanced light intensity at the location of the expanding annular acoustic wave, where there is a positive refractive index perturbation. The annulus serves to trap probe light either when it overfills the entrance of the structure or when it is more tightly focused into the entrance. Either way, the density depression on axis repels probe light from the center of the structure, and a portion of the diverging beam is trapped by the annulus. The results of Figure 4 are for the probe beam overfilling the entrance, but simulation of a more tightly focused probe gives qualitatively similar results. Past the end of the filament (z=48 cm), the annular beam generates a strong interference maximum on axis within a few centimeters, producing the appearance of on-axis guiding if the plane containing the maximum is imaged onto a camera. (The well-known Bessel beam [19] can be considered as an example of a self-interfering annular beam.) The range of time delays for which an apparent on-axis mode is seen depends on the position of the object plane and the diameter of the acoustic annulus, which is set by the sound speed and time delay. Beyond ~1 µs, the annulus has moved sufficiently outward that the interference maximum shifts beyond the imaging system's object plane. The acoustic wave's amplitude, and thus its ability to trap light, also rapidly falls as it propagates outward. These effects create the appearance of an optimum temporal window for light trapping. Non-interferometric probe images for our ~20 cm filaments show on-axis maxima whose behaviour is consistent with this picture.

For a 150ns probe pulse [8], these effects should be present, though temporally smeared out.

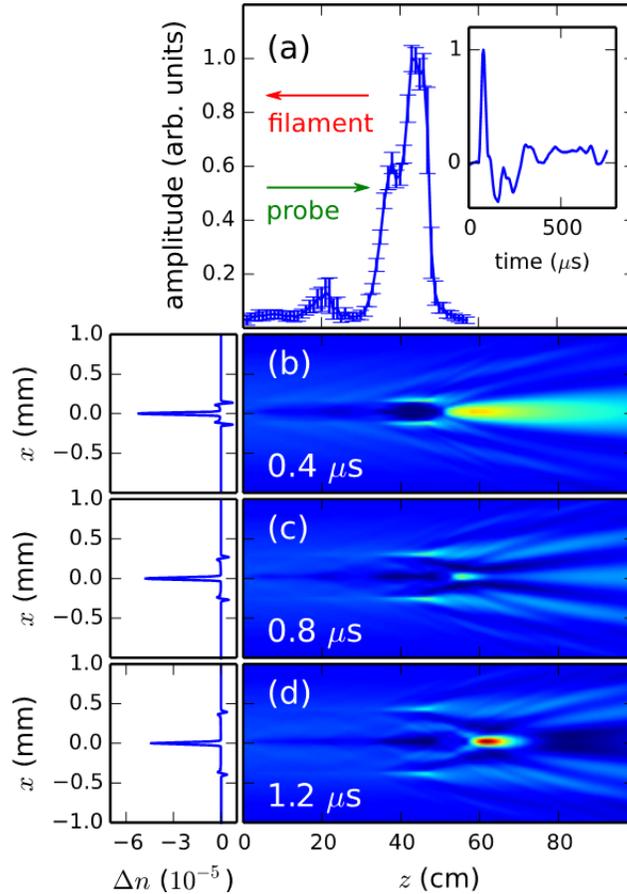

**Figure 4.** Measurement of the axial dependence of energy deposition in a single filament and simulation of the propagation of a probe beam in the post-filament gas density profile. (a) Measured acoustic amplitude along the filament. The filamenting pulse propagates right-to-left. Error bars indicate the standard deviation of the signal amplitude over 100 shots at each position. The inset shows a sample acoustic trace measured by the microphone. (b-d) Beam propagation method simulations of a probe beam propagating left-to-right using hydrocode-generated gas index profiles weighted by the acoustic data of (a). The leftmost panels show index profiles associated with the maximum acoustic amplitude (near $z = 42$ cm) for the post-filament time delays in (b-d).

In conclusion, we have directly measured the sub-microsecond gas dynamics following ultrashort pulse single- and multi-mode filamentation in air, obtaining excellent agreement with hydrodynamic simulations. Accurate measurements depend on minimizing refractive distortion of the probe, which demands a short probe interaction length. For the post-single filament gas response, measurements and simulations show that while there is no positive on-axis refractive index enhancement at any delay, the annular sound wave can trap injected light during a ~ 1 µs window, leading to the appearance of an on-axis interference mode of limited axial extent beyond the end of the filament. An array of filaments, however, can produce an on-axis refractive index enhancement owing to superposition of acoustic waves, and this structure may serve as an optical waveguide, as recently verified [9].

We have also recently shown that an even more effective and much longer-lived waveguide can be generated using the relaxation of a filament array in the thermal diffusion regime [9].

This work is supported by the Air Force Office of Scientific Research, the National Science Foundation, and the US Dept. of Energy. The authors thank R. Birnbaum for technical assistance.

**References**

[1] J. Yu, D. Mondelain, J. Kasparian, E. Salmon, S. Geffroy, C. Favre, V. Boutou, and J.-P. Wolf, Applied Optics 42, 7117 (2003).
[2] Z.-Y. Zheng, J. Zhang, Z.-Q. Hao, Z. Zhang, M. Chen, X. Lu, Z.-H. Wang, and Z.-Y. Wei, Optics Express 13, 10616 (2005).
[3] D. V. Kartashov, A. V. Kirsanov, A. M. Kiselev, A. N. Stepanov, N. N. Bochkarev, Y. N. Ponomarev, and B. A. Tikhomirov, Optics Express 14, 7552 (2006).
[4] B. Clough, J. Liu, and X.-C. Zhang, Optics Letters 35, 3544 (2010).
[5] F. Vidal, D. Comtois, C.-Y. Chien, A. Desparois, B. La Fontaine, T. Johnston, J. Kieffer, H. P. Mercure, H. Pepin, and F. Rizk, IEEE Transactions on Plasma Science 28, 418 (2000).
[6] S. Tzortzakis, B. Prade, M. Franco, A. Mysyrowicz, S. Hller, and P. Mora, Physical Review E 64, 057401 (2001).
[7] Y.-H. Cheng, J. K. Wahlstrand, N. Jhajj, and H. M. Milchberg, Optics Express 21, 4740 (2013).
[8] L. Levi, O. Lahav, I. Kaminer, M. Segev, and O. Cohen, "Long-Lived Waveguides and Sound Wave Generation by Laser Filamentation," arXiv:1307.3588 (2013).
[9] N. Jhajj, E. W. Rosenthal, R. Birnbaum, J. K. Wahlstrand, and H. M. Milchberg, "Demonstration of long-lived high power optical waveguides in air," arXiv:1311.1846 (2013).
[10] N. Jhajj, Y.-H. Cheng, J. K. Wahlstrand, and H. M. Milchberg, Optics Express 21, 28980 (2013).
[11] S. Zahedpour, J. K. Wahlstrand, and H. M. Milchberg, "Quantum control of molecular hydrodynamics," submitted.
[12] K. P. Birch, Journal of the Optical Society of America A 8, 647 (1991).
[13] D. R. Lide, ed. *CRC Handbook of Chemistry and Physics*, 77th ed. (CRC Press, Boca Raton, Fla., 1996).
[14] K. Y. Kim, I. Alexeev, and H. M. Milchberg, Opt. Express 10, 1563 (2002).
[15] L. Liu, C. Wang, Y. Cheng, H. Gao, and W. Liu, Journal of Physics B: Atomic, Molecular and Optical Physics 44, 215404 (2011).
[16] H. Gao, W. Chu, G. Yu, B. Zeng, J. Zhao, Z. Wang, W. Liu, Y. Cheng, and Z. Xu, Optics Express 21, 4612 (2013).
[17] http://lasermatter.umd.edu/HTML%20Files/Publications.htm
[18] M. D. Feit and J. Fleck, Applied Optics 17, 3990 (1978).
[19] J. Fan, E. Parra, and H.M. Milchberg, Phys. Rev. Lett. 84, 3085 (2000).